\documentclass[12pt,preprint]{aastex}
\usepackage{graphicx}

\begin{document}

\title{Detection of Two Carbon-Chain--Rich Cores; CB130-3 and L673-SMM4}
\author{Tomoya HIROTA}
\affil {National Astronomical Observatory of Japan, Osawa 2-21-1, Mitaka, Tokyo 181-8588, Japan; }
\affil {tomoya.hirota@nao.ac.jp}
\author{Takeshi SAKAI}
\affil {Institute of Astronomy, The University of Tokyo, Osawa 2-21-1, Mitaka, Tokyo 181-0015, Japan}
\and
\author{Nami SAKAI, Satoshi YAMAMOTO}
\affil {Department of Physics and Research Center for the Early Universe,}
\affil {The University of Tokyo, Hongo 7-3-1, Bunkyo-ku, Tokyo 113-0033, Japan}

\begin{abstract}
We have found two dense cores, CB130-3 and L673-SMM4, 
where the carbon-chain molecules are extremely abundant relative to NH$_{3}$, 
during a survey observation of radio emission 
lines of CCS, HC$_{3}$N, HC$_{5}$N, and NH$_{3}$ toward dark cloud cores. 
Judging from the low NH$_{3}$/CCS ratios, they are possible candidates for 
"Carbon-Chain--Producing Regions (CCPRs)" recognized 
as chemically young dark cloud cores. The deuterium fractionation ratios 
DNC/HN$^{13}$C in CB130-3 and L673-SMM4 
are found to be 1.28$^{+0.27}_{-0.05}$ and 1.96$^{+0.32}_{-0.01}$, 
respectively, which are comparable to or slightly higher than 
those in CCPRs found previously. 
We suggest that the dense cores of CB130-3 and L673-SMM4 are analogous to 
CCPRs but their chemical evolutionary phase would be 
slightly older than those of the dense cores in the Taurus region. 
\end{abstract}

\keywords{ISM:abundances --- ISM:Individual objects (CB130-3, L673-SMM4) --- ISM:Molecules --- radio lines: ISM}

\section{Introduction}
Since the discovery of carbon-chain molecules, 
their spectral lines have been employed 
as one of the best tracers of dense molecular cloud cores 
\citep[e.g.,][]{suzuki1992}. 
Both abundances and distributions of carbon-chain molecules 
are found to be anti-correlated with 
those of NH$_{3}$ and N$_{2}$H$^{+}$. 
The representative sources observed extensively 
are TMC-1 \citep{hirahara1992} and 
L1544 \citep{ohashi1999} in the Taurus Molecular Cloud. 
The anti-correlation can be interpreted as a consequence of chemical evolution of dark cloud cores, 
because carbon-chain molecules and nitrogen-bearing inorganic molecules 
are expected to be abundant in the early and late stages of chemical evolution, 
respectively \citep{suzuki1992, bergin1997, aikawa2001, aikawa2003, aikawa2005}. 
Thus, dense cores where carbon-chain molecules are abundant 
are thought to be in a very early phase of chemical and dynamical evolution. 

According to the pioneering survey of the CCS lines by \citet{suzuki1992}, 
four dense cores were identified where extraordinarily intense 
spectra of carbon-chain molecules were detected in spite of very weak NH$_{3}$ lines. 
These four sources, L1495B, L1521B, L1521E, and the cyanopolyyne peak of TMC-1, 
are named as "Carbon-Chain--Producing Regions" (hereafter called CCPR). 
Until our recent discovery of another candidate L492 in the Aquila rift region 
\citep{hirota2006, hirota2009}, CCPRs were identified only in the Taurus Molecular Cloud 
\citep{hirota2002, hirota2004, tafalla2004}. 
Based on the previous survey observations, the fraction of CCPRs among all dense cores is 
found to be about 10\% \citep{suzuki1992, hirota2009}. 
This means that the timescale staying in the CCPR phase is expected to be 
shorter than a typical lifetime of dense cores by a factor of 10. 
This is consistent with the idea that the evolutionary stage of CCPRs is relatively 
younger than other typical dark cloud cores. 
Thus, CCPRs would be the best targets to investigate early stages for 
the gas-phase chemistry under extremely low degree of depletion, as well as 
dynamical state of dense cores just after dense core formation. 

However, CCPRs have rarely been detected outside the Taurus region. 
According to \citet{hirota2009}, there could be possible chemical differentiation 
among molecular clouds including the Taurus, Ophiuchus, and Aquila regions. 
For instance, detection rates of carbon-chain molecules are found to be 
systematically higher in the Taurus region than those in Ophiuchus. 
This implies that the average duration time of the starless core phase is different 
from cloud to cloud. However, we cannot rule out the possibility that the higher 
detection rate of CCPRs is due to some specific nature in the Taurus region. 
In order to examine this possibility, a search for further CCPR candidates other 
than in the Taurus region is useful. 

With this motivation, 
we carried out a survey observation of the CCS, HC$_{3}$N, HC$_{5}$N, and 
NH$_{3}$ lines toward more than 100 dark cloud cores \citep{hirota2009, hirota2011}. 
As a result, we detected two dense cores, CB130-3 and L673-SMM4, 
which show remarkably intense spectra of carbon-chain molecules. 
CB130-3 is identified as an optical dark cloud core by \citet{lee1999} in the 
Aquila rift region at the distance of 200~pc. CB130-3 resides in the globular filament 
GF4 \citep{schneider1979}, where another CCPR, L492, is also located 
as shown in Figure \ref{fig-gf4}. 
L673-SMM4 is one of the dust continuum peaks \citep{visser2002} in the dark cloud 
L673 located in the Cloud~B region at the distance of 300~pc \citep{dame1987}. 
In this paper, we present results of our intensive observations of newly found CCPR 
candidates. Detail for the survey observation will be described in the 
forthcoming paper \citep{hirota2011}. 

\section{Observations \label{sec-observations}}

We first conducted the survey observation with the 45~m radio telescope 
at Nobeyama Radio Observatory (NRO) in 2008 May and 2009 April. 
The CCS ($J_{N}$=4$_{3}$-3$_{2}$), 
HC$_{3}$N ($J$=5-4), and HC$_{5}$N ($J$=17-16) lines 
in the 45 GHz region were observed simultaneously with 
a Superconductor-Insulator-Superconductor (SIS) mixer receiver, 
whose system temperature was 190-260~K. 
The main-beam efficiency ($\eta_{mb}$) and the beam size (HPBW) were 0.7 
and 37\arcsec, respectively. 
During the survey observation, we serendipitously found strong emissions 
of the CCS, HC$_{3}$N, and HC$_{5}$N lines toward 
CB130-3 and L673-SMM4. Subsequently, we made mapping observations of 
the lines in these two sources with a grid spacing of 40\arcsec. 
The source coordinates are listed in Table \ref{tab-source}. 
In addition, we observed the molecular lines in the 72-93~GHz band 
as listed in Table \ref{tab-obsline} toward the CCS peak in each core. 
One of the SIS mixer receivers equipped with the NRO 45~m telescope was used 
for the observations of the 72-93~GHz lines. 
The system temperatures ranged from 180 to 400~K, depending on the frequencies. 
The main-beam efficiency and the beam size were approximately 
0.44 and 20\arcsec, respectively, although these values depend 
on the observed frequencies. 
For all of the frequency bands, acousto-optical radio spectrometers with 
the frequency resolution of 37 kHz were used as the back end. 
Observations were performed in the position-switching mode 
with the off position of 10\arcmin \ toward the azimuth direction. 
The antenna temperatures were calibrated by using the chopper-wheel 
method \citep{ulich1976}. 
Pointing was checked every two hours by observing the nearby SiO maser sources, 
V446~Oph and RR~Aql, for observations of CB130-3 and L673-SMM4, respectively, 
and the pointing accuracy was kept to be better than 5\arcsec rms. 
The observed data were analyzed with the software package NEWSTAR developed by NRO. 

The NH$_{3}$ lines in the 23~GHz band were also observed with the 100~m 
radio telescope of Max Planck Institute for Radio Astronomy (MPIfR) in 
Effelsberg. The observations were carried out in 2010 February. 
A dual polarization high electron mobility transistor (HEMT) receiver 
was employed with the system temperature of 
50-90~K. The mapping observations were made with the grid spacing of 
40\arcsec \ in the frequency switching mode. 
The fast Fourier transform (FFT) 
spectrometer was used for the back end, of which 
spectral resolution was set to be 10~kHz. The pointing and focusing were 
checked by observing a continuum emission from the quasar 1741-038. 
The flux calibration was conducted by observing NGC~7027, whose 
flux density is calculated to be 5.64~Jy with a program provided by the observatory 
\citep[e.g.,][]{ott1994}. 
The data were analyzed with the CLASS software package. 

\section{Results}

\subsection{Molecular Abundances}

\begin{figure*} \begin{center}
\includegraphics[width=15cm]{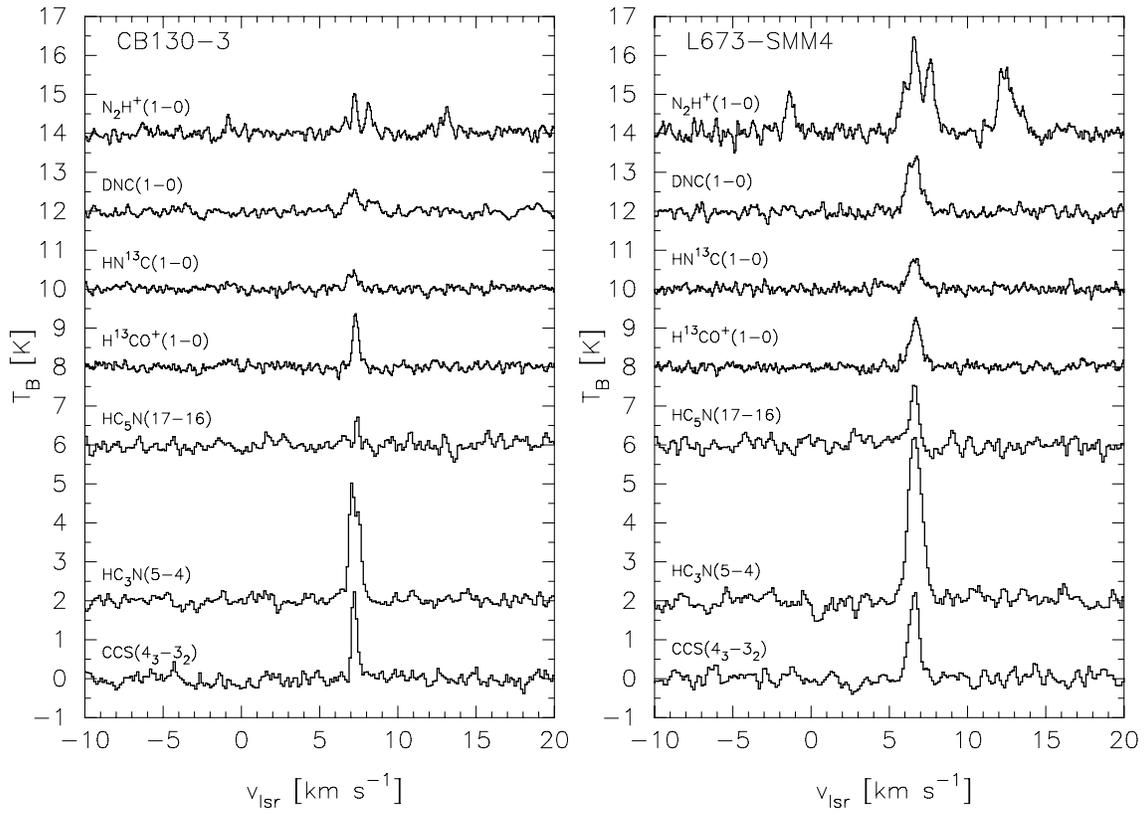}
\caption{Line profiles observed toward CB130-3 and L673-SMM4. \label{fig-spectra}  }
\end{center} \end{figure*}

Observed spectra are shown in Figures \ref{fig-spectra} and \ref{fig-nh3spectra}. 
The intensities of the CCS, HC$_{3}$N, and HC$_{5}$N lines toward CB130-3 and 
L673-SMM4 are found to be remarkably strong, 
and are almost comparable to those toward the previously known 
CCPRs \citep{hirota2002, hirota2004, hirota2006}. 
The peak brightness temperature, line width, 
and LSR velocity are derived by fitting the Gaussian profile to each 
spectrum. The results are summarized in Table \ref{tab-obsline}. 
The line widths and LSR velocities for HC$_{3}$N, DNC, and HN$^{13}$C 
are slightly different from one another due to their unresolved hyperfine structures. 

Using these line parameters, column densities of CCS, HC$_{3}$N, and HC$_{5}$N
were calculated by the method which is consistent with that by \citet{suzuki1992} 
and the previous paper \citep{hirota2009}. 
The dipole moments of CCS, HC$_{3}$N, and HC$_{5}$N were assumed 
to be 2.81, 3.72, and 4.33 D, respectively 
\citep{murakami1990, lafferty1978, alexander1976}. 
The LTE condition was assumed, and the excitation temperature 
was fixed to 6.0~K for CCS, as in the case of other CCPRs 
\citep{suzuki1992, hirota2001}. 
Note that the column density of CCS could not be calculated by 
assuming the excitation temperature of 5.0~K due to the 
high brightness temperature for CB130-3. 
For the HC$_{3}$N lines in CB130-3 and L673-SMM4, 
we could not detect the satellite hyperfine components ($\Delta F=0$) 
to determine the optical depths and the excitation temperatures. 
Thus, we simply adopted the excitation 
temperature of 6.5~K for CB130-3 as employed in the previous 
papers for consistency \citep[e.g.,][]{suzuki1992, hirota2009}. 
On the other hand, the higher excitation temperature of 
7.5~K was required to calculate the column density of HC$_{3}$N 
in L673-SMM4 due to its exceptionally high brightness temperature. 
Under these assumptions, the upper limit to the optical depth of 
the HC$_{3}$N lines is estimated to be $<$3.7 and $<$2.7 for 
CB130-3 and L673-SMM4, respectively (1$\sigma$). 
The excitation temperature of HC$_{5}$N is assumed to be 6.5~K 
as in the previous studies \citep{suzuki1992, hirota2009}. 

For the NH$_{3}$ (1, 1) line, we derived the excitation temperature 
and total optical depth based on the hyperfine fitting program 
in the CLASS software package. 
As a result, we could determine both parameters 
toward L673-SMM4, as summarized 
in Table \ref{tab-obsnh3}. In contrast, the hyperfine fitting yields the 
excitation temperature of the NH$_{3}$ line toward CB130-3 
to be 4.0$\pm$1.2~K, which seems too low in comparison with those in the typical NH$_{3}$ 
cores \citep{suzuki1992} and other sources in the present study (Table \ref{tab-obsnh3}). 
Because of the low signal to noise ratio for the hyperfine components of the NH$_{3}$ 
lines toward CB130-3, the fitting result would not be reliable. 
Thus, we fitted 
the single Gaussian to the brightest hyperfine component of 
the NH$_{3}$ spectrum, and simply adopted the excitation 
temperature of 6.5~K for CB130-3 as in the previous papers for consistency 
\citep{hirota2002, hirota2004, hirota2006}. 
The line width for CB130-3 derived from the single Gaussian fitting 
is apparently broadened due to unresolved hyperfine components. 
We calculated the column density of NH$_{3}$ under the assumption of the rotation 
temperature of 10~K and the ortho-to-para ratio of 1 (the statistical equilibrium value). 
The dipole moment employed is 1.46 D \citep{cohen1974}. 

The results of the column density calculations 
are summarized in Table \ref{tab-collte}. 
We estimated the uncertainties in the derived column densities as follows. 
For CCS, HC$_{3}$N, and HC$_{5}$N, the derived column densities differ 
within 30\% except for HC$_{3}$N in L673-SMM4, 
when we change the excitation temperature of 0.5~K, as listed in Table \ref{tab-collte}. 
In the case of HC$_{3}$N in L673-SMM4, the column density differs by 
a factor of 1.8, because of the higher optical depth. 
For NH$_{3}$, we estimated the uncertainties in their column densities 
by considering the errors of the excitation temperatures in 
the hyperfine fitting, as summarized in Table \ref{tab-obsnh3}. 
For CB130-3, we changed the excitation temperature of 2~K to 
evaluate the possible range of the column density. 
As a result, the NH$_{3}$/CCS ratio, 
which is employed as a good indicator of chemical evolutionary phase, 
would contain uncertainties of about 50\%, as summarized in Table \ref{tab-dhratio}. 
Because the assumed excitation temperatures are critical 
to estimate accurate abundances, further refinement 
should be made based on multi-transition observations. 

The total optical depths for the N$_{2}$H$^{+}$ ($J$=1-0) lines are derived by 
the hyperfine fitting to be 2.3$\pm$0.3 and 6.6$\pm$1.1 
for CB130-3 and L673-SMM4, respectively. 
The LSR velocities and line widths are also determined, as shown 
in Table \ref{tab-obsline}. 
Using these results, we obtained the column densities of N$_{2}$H$^{+}$ 
by assuming the LTE condition as summarized in Table \ref{tab-collte}. 
When we attempted to determine both the excitation temperature and 
optical depth simultaneously, the least-squares fitting program did not 
converge with an appropriate solution. 
Thus, we fixed the excitation temperature of N$_{2}$H$^{+}$ to be 5.0~K 
to fit all of the hyperfine components \citep{benson1998}. 
The uncertainty in the column density is estimated by changing the assumed 
excitation temperature by 0.5~K. 
Note that the brightness temperature of the main hyperfine component in 
L673-SMM4 is underestimated in comparison with the observed spectrum 
(see Table \ref{tab-obsline} and Figure \ref{fig-spectra}). This indicates 
higher excitation temperature and lower optical depth in L673-SMM4. 
If we employed the excitation temperature of 10~K, the fitting result 
was improved. In this case, the total optical depth was derived to be 1.40, 
and the derived column density decreased by a factor of 1.5, 
8.0$\times10^{12}$~cm$^{-2}$. 

For DNC, HN$^{13}$C, and H$^{13}$CO$^{+}$, 
we derived the column densities using the large velocity gradient (LVG) model 
\citep{goldreich1974}, as adopted in \citet{hirota2001, hirota2003}. 
Details of the analysis are described elsewhere \citep{hirota2006}. 
Because we observed only the $J$=1-0 lines, the H$_{2}$ density 
and the kinetic temperature were fixed to be 1$\times$10$^{5}$~cm$^{-3}$ 
and 10~K, respectively. 
To evaluate the uncertainty in the derived column density, 
we made the LVG analyses by varying the H$_{2}$ density from 
5$\times$10$^{4}$ to 5$\times$10$^{5}$~cm$^{-3}$. 
As a result, we confirmed that the DNC/HN$^{13}$C ratios change 
only by a factor of 1.2 within the above H$_{2}$ density range, 
as shown in Table \ref{tab-dhratio}. 

Table \ref{tab-dhratio} lists the DNC/HN$^{13}$C and NH$_{3}$/CCS 
ratios of a few representative sources in addition to those of CB130-3 and 
L673-SMM4. These two ratios are known to be good indicators of 
chemical evolutionary stages \citep{hirota2006}. We show a plot of these ratios 
in Figure \ref{fig-plots}, as prepared by \citet{hirota2001}. 
Both the NH$_{3}$/CCS and deuterium fractionation ratios in 
CB130-3 and L673-SMM4 are systematically lower than those in typical 
dark cloud cores. 
In our previous paper \citep{hirota2009}, we classified CCPRs as 
dense cores with the NH$_{3}$/CCS ratios lower than 10. 
The NH$_{3}$/CCS ratios in CB130-3 and L673-SMM4 are 
slightly higher than the above threshold, indicating that these two sources 
are close to the evolved prestellar cores such as L1544 (see Table \ref{tab-dhratio}). 
The NH$_{3}$/CCS ratio in L492, which is reevaluated in this paper on the basis 
of the follow-up observation of NH$_{3}$ (Appendix), is almost 
comparable to those of CB130-3 and L673-SMM4. 
Nevertheless, it should be emphasized that 
the NH$_{3}$/CCS ratios in these cores are still 
lower than those in typical dark cloud cores. 

Note that we here employed the NH$_{3}$ data obtained with the 
MPIfR 100~m telescope, which has higher spatial resolution than 
the NRO~45~m telescope, and hence, beam dilution effects 
would be less significant. In other words, the NH$_{3}$/CCS ratios 
obtained with the 45~m telescope are possibly underestimated 
in comparison with those with the 100~m telescope. 
In fact, we confirmed that the brightness temperatures of the NH$_{3}$ line 
toward the reference positions (0\arcsec, 0\arcsec) of CB130-3 and 
L673-SMM4 observed with the NRO 45~m telescope were $\sim$3 times weaker 
than those with the MPIfR 100~m telescope. From this result, 
the equivalent FWHM source size of the NH$_{3}$ cores is estimated to be 
$\sim$30\arcsec, which is consistent with the NH$_{3}$ core size derived 
in the mapping observation of this study (Sections 3.2 and 3.3; 
Figures \ref{fig-mapcb130} and \ref{fig-mapl673}). 

On the basis of the relatively lower NH$_{3}$/CCS and DNC/HN$^{13}$C ratios 
in CB130-3 and L673-SMM4, we conclude that CB130-3 and L673-SMM4 would be 
chemically younger than typical dense cores. 
This means that they are new candidates for CCPRs outside the Taurus region. 
On the other hand, these ratios in CB130-3 and L673-SMM4 
along with those in L492 are slightly higher than those in CCPRs in the Taurus region, 
as summarized in Table \ref{tab-dhratio} and Figure \ref{fig-plots}. 
Thus, it is likely that CB130-3, L492, and L673-SMM4 are slightly more evolved 
than other CCPRs in the Taurus region. 

\subsection{Molecular Distributions in CB130-3}

\begin{figure*} \begin{center}
\includegraphics[width=13.0cm]{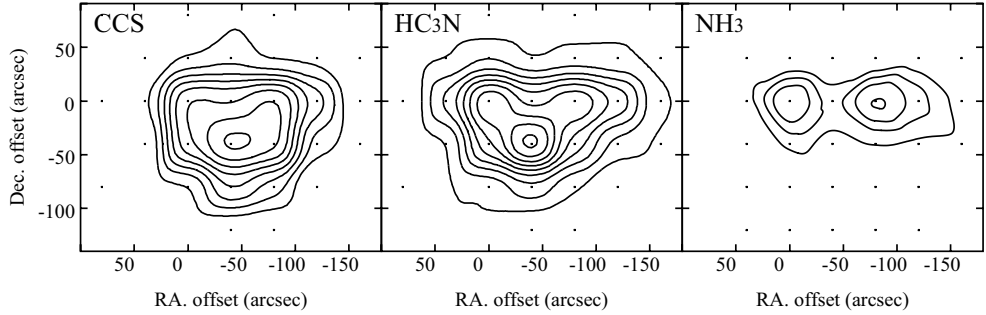}
\caption{Integrated intensity maps of CB130-3. 
The reference position is listed in Table \ref{tab-source}. 
The velocity range of integration, lowest contour levels, and 
contour intervals are 
(6.4-7.8~km~s$^{-1}$, 0.23~K~km~s$^{-1}$, 0.11~K~km~s$^{-1}$) for CCS, 
(6.4-8.0~km~s$^{-1}$, 0.23~K~km~s$^{-1}$, 0.23~K~km~s$^{-1}$) for HC$_{3}$N, 
and (6.5-8.5~km~s$^{-1}$, 0.60~K~km~s$^{-1}$, 0.30~K~km~s$^{-1}$) for 
the main hyperfine component of NH$_{3}$ (1,1). 
\label{fig-mapcb130} }
\end{center} \end{figure*}

We obtained the integrated intensity maps of the CCS, HC$_{3}$N, and NH$_{3}$ lines 
toward CB130-3 as shown in Figure \ref{fig-mapcb130}, 
while the HC$_{5}$N lines are too weak to prepare the map. 
Both the HC$_{3}$N and CCS distributions show a centrally condensed structure 
without any signature of a dip or a hole caused by molecular depletion, although 
the distribution of HC$_{3}$N is slightly more extended along the east-west direction 
than the CCS distribution.

In contrast, the NH$_{3}$ distribution shows a double-peaked structure 
elongating along the east-west direction. 
The size of the substructure seen in the NH$_{3}$ map is 
smaller than the CCS and HC$_{3}$N distributions. 
Since the beam size of the CCS and HC$_{3}$N observations, 40\arcsec, 
is almost comparable to that of the NH$_{3}$ observation, this is not due to 
the spatial resolution. Considering that the observed lines have similar 
critical densities, this seems to reflect the chemical differentiation. 

Such anti-correlation between the distributions of NH$_{3}$ and carbon-chain 
molecules like CCS and HC$_{3}$N 
is often found in dense cores \citep{hirahara1992, ohashi1999}. 
In the starless core L1544, NH$_{3}$ shows a centrally condensed distribution, 
whereas CCS has a doughnut-like distribution around it \citep{ohashi1999, tafalla2002}. 
This means that 
the central part is more chemically evolved than the outer part. 
However, the CB130-3 case shows an opposite trend; carbon-chain molecules 
are centrally condensed while NH$_{3}$ is distributed outside the carbon-chain 
molecules. 
One possibility is that two compact clumps are being formed 
in the dense core of CB130-3. Chemically evolved clumps 
traced by the NH$_{3}$ lines would just be formed by the fragmentation of 
the parent core traced by the CCS and HC$_{3}$N lines. However, we could find 
no specific velocity/spatial structures suggesting such small scale clumps 
in the CCS map. 
Because of the lack of high-resolution 
dust continuum observations, the density profile 
and dynamical properties of this core are still unclear. 
It would be crucial to observe the millimeter and submillimeter 
dust continuum emissions in order to understand the different distributions 
of NH$_{3}$ and CCS in CB130-3. 

The line width of the CCPRs is very interesting in relation to their dynamical state. 
If the cores are self-gravitating in the CCPRs, the large line width may be expected. 
Turbulent motions in the surrounding medium could still be retained in the young 
core, or the accretion motion of the chemically young material onto dense cores may 
contribute to the line width. However, we found that the line widths of CB130-3 are 
not significantly different. 
The observed line widths for CB130-3 are also similar to those for other CCPRs 
\citep{hirota2002, hirota2004, hirota2006}.  
An asymmetric line profile indicating infalling motion 
has not been detected toward CB130-3 \citep{lee1999inf}. 
It is then suggested that CB130-3 is in dynamically less evolved phase 
than in other prestellar cores such as L1544, L1498 and another 
CCPR L492 \citep{hirota2006}. 
We need more intensive observations to detect the dynamical signatures expected for 
young clouds. 

\subsection{Molecular Distributions in L673-SMM4}

\begin{figure*} \begin{center}
\includegraphics[width=17.0cm]{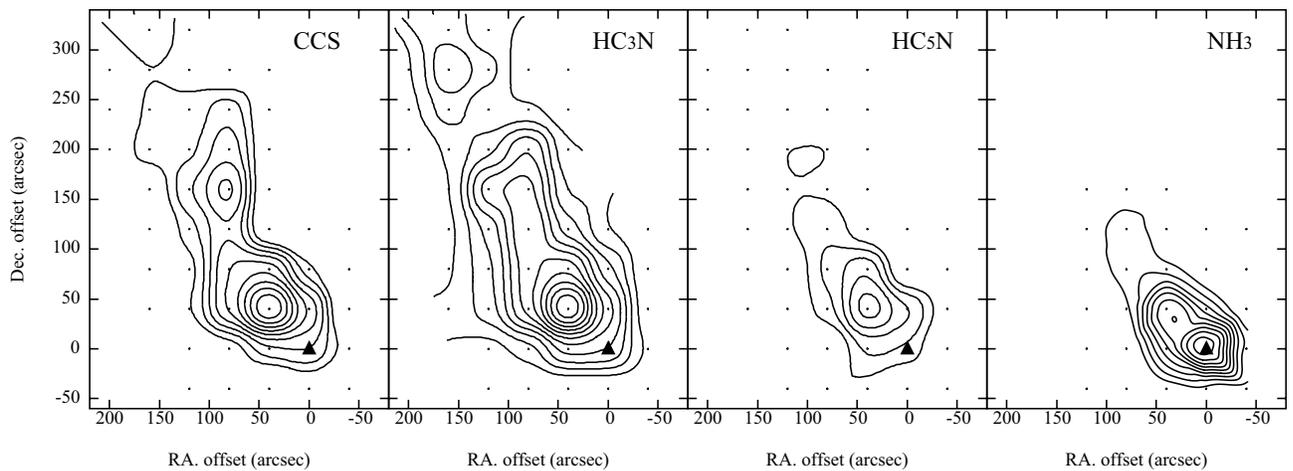}
\caption{Integrated intensity maps of L673-SMM4. 
The reference position is listed in Table \ref{tab-source}. 
The velocity range of integration, lowest contour levels, and 
contour intervals are 
(5.8-7.4~km~s$^{-1}$, 0.29~K~km~s$^{-1}$, 0.14~K~km~s$^{-1}$) for CCS, 
(5.6-7.8~km~s$^{-1}$, 0.34~K~km~s$^{-1}$, 0.34~K~km~s$^{-1}$) for HC$_{3}$N, 
(5.8-7.4~km~s$^{-1}$, 0.29~K~km~s$^{-1}$, 0.14~K~km~s$^{-1}$) for HC$_{5}$N, 
and (6.0-7.5~km~s$^{-1}$, 0.60~K~km~s$^{-1}$, 0.30~K~km~s$^{-1}$) for 
the main hyperfine component of NH$_{3}$ (1,1). 
A triangle in each map represents the submillimeter continuum 
source SMM4 \citep{visser2002}. 
\label{fig-mapl673} }
\end{center} \end{figure*}

For L673-SMM4, the CCS, HC$_{3}$N, NH$_{3}$, and even HC$_{5}$N 
lines are sufficiently intense to map the whole region of the core. 
The obtained maps are shown in Figure \ref{fig-mapl673}. 
The molecular distributions show an elongated ridge-like structure along the 
north-south direction. The HC$_{5}$N line only traces the southern part 
of the ridge. The peak position for CCS, HC$_{3}$N, and HC$_{5}$N is 
significantly shifted from the dust continuum peak SMM4 \citep{visser2002} 
toward northeast by (40\arcsec, 40\arcsec). 
One can see the sub-peak at the (80\arcsec, 160\arcsec) position 
offset from the dust continuum peak in the CCS map. 
This sub-peak is also evident both in the HC$_{3}$N and HC$_{5}$N 
maps, in which an elongated structure and a weak emission peak, 
respectively, are detected. 
In addition, there is an indication of extended structure over 
the northern edge of the CCS and HC$_{3}$N maps. 

\begin{figure} \begin{center}
\includegraphics[width=7.5cm]{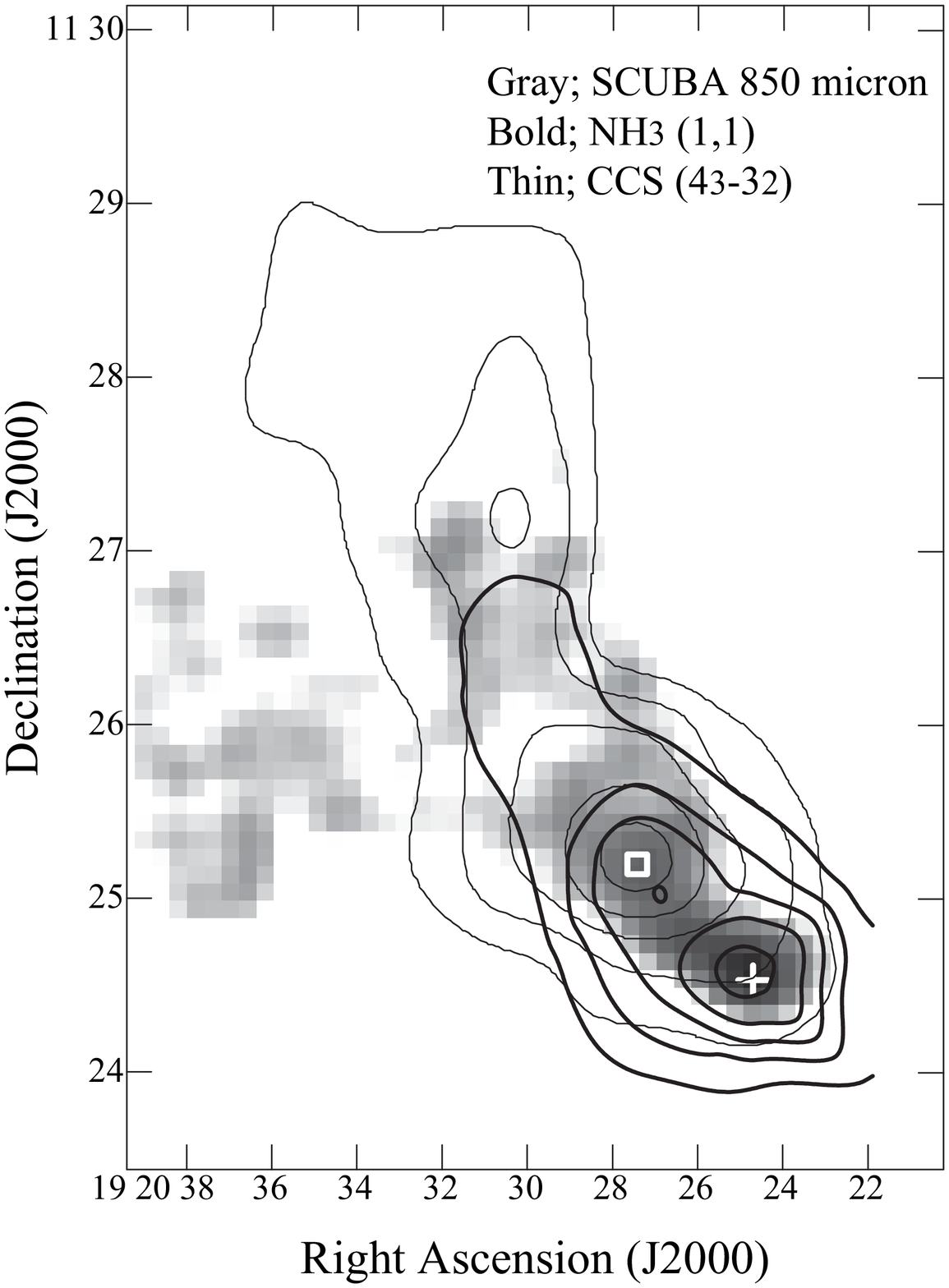}
\caption{Integrated intensity maps of L673-SMM4 superposed 
on the dust continuum map taken from the SCUBA Legacy Catalog 
\citep{difrancesco2008}. 
The gray scale represents the dust continuum emission at 
850~$\mu$m. The bold and thin contours show the 
integrated intensity of the CCS and NH$_{3}$ lines, respectively. 
The lowest contour levels and contour intervals are 
(0.29~K~km~s$^{-1}$, 0.29~K~km~s$^{-1}$) for CCS and 
(0.60~K~km~s$^{-1}$, 0.60~K~km~s$^{-1}$) for 
the main hyperfine component of NH$_{3}$ (1,1). 
A cross and a square indicate the positions of the NH$_{3}$ and CCS peaks, 
respectively. 
\label{fig-scuba} }
\end{center} \end{figure}

On the other hand, the NH$_{3}$ peak is in good agreement with the 
dust continuum peak, as shown in Figure \ref{fig-scuba}. 
In addition, we can marginally see the second peak 
which coincides with the CCS peak position. 
At the second NH$_{3}$ peak (i.e., the CCS peak), there is 
a signature of a sub-peak in the dust continuum map. 
The dust continuum emission also seems to be around the northern CCS peak. 
Therefore, the NH$_{3}$ map well traces the density structure, 
while the CCS emission shows a possible evidence of depletion at 
the density peak, or preferentially traces a lower density region. 

Although the NH$_{3}$ emission extends northeastward beyond the CCS, HC$_{3}$N, 
and HC$_{5}$N peaks, its emitting region is smaller than 
those of carbon-chain molecules. 
Thus, the NH$_{3}$/CCS ratio shows systematic gradient along 
the ridge; the NH$_{3}$/CCS ratio is higher in the southern part 
around the dust continuum peak and decreases toward the northern part. 
As demonstrated in the TMC-1 ridge \citep{hirahara1992}, the chemical abundance 
gradient is explained in terms of 
a sequential evolution of the filamentary cloud. 
If this is the case, the molecular ridge extending north from the L673-SMM4 
would contract first at the southern part of the cloud, 
and then the contraction would be propagating toward the northern part. 
We note that the observed line profiles toward the 
CCS peak in L673-SMM4 are systematically broader than 
those in CB130-3 and typical dense cores. 
This might reflect possible effects of star formation 
such as infalling and/or outflow motions, 
although the infalling line asymmetry has not been searched for L673-SMM4. 
Previously, L673-SMM4 is reported to be a starless core \citep{visser2002}, 
while a faint protostar candidate has been identified toward the submillimeter dust 
continuum peak SMM4 based on the mid-infrared observations with 
the Spitzer Space Telescope \citep{tsitali2010}. 
The broader line widths in L673-SMM4 could be affected 
by the newly formed protostar in the vicinity of the submillimeter peak position. 
If so, L673-SMM4 is regarded as a chemically young core with a signature 
of star forming activities. 

\section{Discussions}

We have carried out molecular line observations of newly found 
dark cloud cores, CB130-3 and L673-SMM4, 
showing remarkably intense spectra of the carbon-chain 
molecules such as CCS, HC$_{3}$N, and HC$_{5}$N. We have also 
observed the NH$_{3}$ lines and other fundamental molecular 
lines toward these two cores. Judging from the NH$_{3}$/CCS ratios 
and DNC/HN$^{13}$C ratios, CB130-3 and L673-SMM4 are analogous to 
the chemically young dark cloud cores called CCPRs 
named by \citet{suzuki1992} but are slightly evolved than 
those found in the Taurus region. 

\begin{figure} \begin{center}
\includegraphics[width=7.5cm]{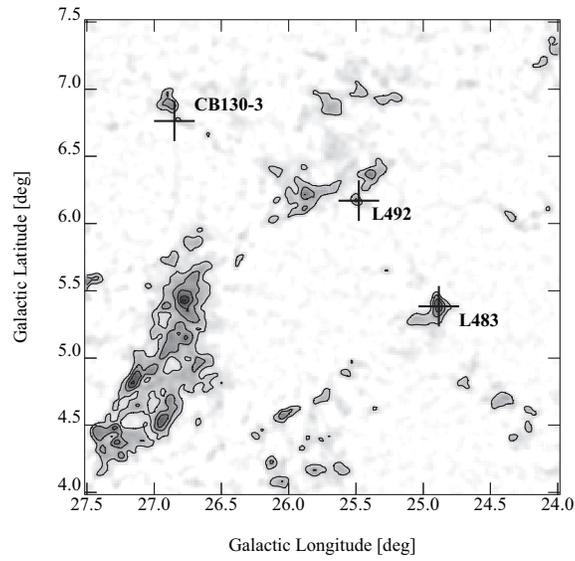}
\caption{Distribution of dark clouds in the GF4 region \citep{schneider1979}. 
Three dense cores, CB130-3, L492, and L483, are 
aligned from the top-left (north) to bottom-right (south) direction. 
The map shows the visual extinction $A_{V}$ composed from the 2MASS extinction 
data \citep{dobashi2011}. The contour levels are $A_{V}$=2, 4, 6, 8, and 10 mag. 
\label{fig-gf4} }
\end{center} \end{figure}

It should be noted that one of the newly found CCPR candidates, CB130-3, 
is located close to other dense cores rich in carbon-chain molecules, 
L492 \citep{hirota2006, hirota2009} and L483 \citep{hirota2010}. 
All of these cores belong to an optical dark cloud GF4 \citep{schneider1979} 
as shown in Figure \ref{fig-gf4}. We speculate that 
chemical properties in GF4 in the Aquila rift region are similar to those 
in the Taurus region. 

According to the chemical model calculations \citep{suzuki1992}, 
carbon-chain molecules are expected to be abundant 
in lower extinction and less dense regions where 
the atomic carbon has not been locked into CO. 
Thus, it is likely that the CCPRs could be lower extinction and less dense 
molecular gas clumps formed in the regions in between translucent and 
dense molecular clouds. 
In fact, some of the CCPRs are identified as relatively isolated small Bok 
globules (e.g., L492 and CB130-3 in the Aquila region) unlike the dense 
molecular cloud such as the Ophiuchus region. 
However, the CCPRs found in the Taurus and Aquila regions have almost similar 
H$_{2}$ densities to other typical dark cloud cores according to the 
statistical equilibrium calculations for some molecular lines 
\citep{hirota2002, hirota2004, hirota2006, aikawa2005}. 
Although we cannot directly compare the H$_{2}$ 
densities in CB130-3 and L673-SMM4, intense spectra of high-density 
tracers such as H$^{13}$CO$^{+}$ imply the H$_{2}$ densities of an order of 
$>$10$^{5}$~cm$^{-1}$. 
Therefore, the CCPRs themselves are not just lower extinction and less dense 
gas clumps but rather chemically younger dense cores compared with 
other prestellar cores such as L1544 and L1498. 
Further observations of these regions including less-dense peripheries around 
dense cores will be necessary to explore a role of a surrounding environment in 
formation of dense cores including CCPRs. 

Existence of CCPRs outside the Taurus region indicates that the origin of 
CCPRs is not ascribed to the regional specialty of the Taurus cloud. 
Hence, our 
results may strengthen the argument that the evolutionary timescale 
of dense cores is different from cloud to cloud, 
as proposed by \citet{hirota2009} and \citet{sakai2009}. 
We recently found another type of carbon-chain rich core in the Lupus 
Molecular Cloud named Lupus-1A \citep{sakai2010, shiino2011}. 
This source shows 
extraordinarily intense spectra of C$_{4}$H and longer carbon-chain molecules, 
while that of CCS is not as bright as in the known CCPRs 
including CB130-3 and L673-SMM4. All of these sources will be good 
targets to reveal chemical differentiation among the large-scale molecular cloud 
complexes. In addition, they can be useful and unique sources to investigate 
formation and evolution of carbon-chain molecules in dark cloud cores in 
the chemically young evolutionary phase. 
They will shed light on the initial state of chemical and dynamical 
evolution of dark cloud cores by the high-resolution observations with 
Atacama Large Millimeter/Submillimeter Array (ALMA).  

\acknowledgements

The 45~m radio telescope is operated by Nobeyama Radio Observatory, 
a branch of National Astronomical Observatory of Japan (NAOJ). 
We are grateful to all the staff of Nobeyama Radio Observatory of NAOJ 
and Effelsberg Observatory of MPIfR for their assistance in observations. 
TH and NS thank to the Inoue Foundation for Science  
for the financial support (Research Aid of Inoue Foundation for Science). 
This study is partly supported by Grant-in-Aid from 
The Ministry of Education, Culture, Sports, Science and Technology of Japan 
(No. 21224002 and 21740132). 

{\it Facilities:} \facility{No:45m,Effelsberg}.

{}

\begin{deluxetable}{lll}
\tabletypesize{\scriptsize}
\tablewidth{0pt}
\tablecaption{Source coordinate \label{tab-source}}
\tablehead{
\colhead{Source} & \colhead{$\alpha$(J2000)} & \colhead{$\delta$(J2000)}  
\vspace{2mm} \\
\colhead{Name} & \colhead{(  h  m  s)} & \colhead{ (  $^{\circ}$ \hspace{0.3em} \arcmin \hspace{0.3em} \arcsec)}
}
\startdata
CB130-3        & 18:16:17.9 & -02:16:41  \\
L673-SMM4      & 19:20:24.6 & +11:24:34 \\
\enddata
\end{deluxetable}

\begin{deluxetable}{lllrllllll}
\rotate
\tabletypesize{\scriptsize}
\tablewidth{0pt}
\tablecaption{List of molecular lines observed with the NRO 45~m telescope
\label{tab-obsline}}
\tablehead{
\colhead{Source Name} & 
  \colhead{Line} & 
  \colhead{$\nu$ (MHz)} & \colhead{$S_{ul}$\tablenotemark{a}} &
  \colhead{$\mu$ (D)\tablenotemark{b}} & \colhead{Reference}  &
  \colhead{$T_{B}$ (K)} & \colhead{$v_{lsr}$ (km s$^{-1}$)} &
  \colhead{$\Delta v$ (km s$^{-1}$)} &  \colhead{$T_{rms}$ (K)} }
\startdata
CB130-3    & CCS($J_{N}$=4$_{3}$-3$_{2}$)             & 45379.033 &  3.97  & 2.81  & 1,2  & 2.26(33) &  7.22(3)  &  0.41(7)    & 0.16   \\
           & HC$_{3}$N($J$=5-4)                       & 45490.316 &  5.00  & 3.72  & 3    & 2.83(24) &  7.22(4)  &  0.81(8)    & 0.13   \\
           & HC$_{5}$N($J$=17-16)                     & 45264.721 & 17.00  & 4.33  & 4    & 0.83(47) &  7.40(8)  &  0.24(17)   & 0.13   \\
           & H$^{13}$CO$^{+}$($J$=1-0)                & 86754.330 &  1.00  & 4.07  & 5    & 1.35(16) &  7.31(3)  &  0.44(6)    & 0.08   \\
           & DNC($J$=1-0)                             & 76305.717 &  1.00  & 3.05  & 6    & 0.48(12) &  7.17(12) &  0.94(27)   & 0.09   \\
           & HN$^{13}$C($J$=1-0)                      & 87090.859 &  1.00  & 3.05  & 6    & 0.41(11) &  7.11(10) &  0.72(22)   & 0.08   \\
           & N$_{2}$H$^{+}$($J$=1-0)\tablenotemark{c} & 93173.777 &  1.00  & 3.40  & 7    & 0.91     &  7.24(4)  &  0.36(5)    & 0.11   \\
L673-SMM4  & CCS($J_{N}$=4$_{3}$-3$_{2}$)             & 45379.033 &  3.97  & 2.81  & 1,2  & 2.19(26) &  6.59(5)  &  0.76(11)   & 0.19   \\
           & HC$_{3}$N($J$=5-4)                       & 45490.316 &  5.00  & 3.72  & 3    & 4.24(27) &  6.65(3)  &  1.02(8)    & 0.14   \\
           & HC$_{5}$N($J$=17-16)                     & 45264.721 & 17.00  & 4.33  & 4    & 1.51(31) &  6.59(7)  &  0.62(15)   & 0.17   \\
           & H$^{13}$CO$^{+}$($J$=1-0)                & 86754.330 &  1.00  & 4.07  & 5    & 1.15(10) &  6.68(4)  &  0.90(9)    & 0.09   \\
           & DNC($J$=1-0)                             & 76305.717 &  1.00  & 3.05  & 6    & 1.36(13) &  6.57(5)  &  1.02(11)   & 0.11   \\
           & HN$^{13}$C($J$=1-0)                      & 87090.859 &  1.00  & 3.05  & 6    & 0.77(11) &  6.60(6)  &  0.86(14)   & 0.09   \\
           & N$_{2}$H$^{+}$($J$=1-0)\tablenotemark{c} & 93173.777 &  1.00  & 3.40  & 7    & 1.65     &  6.65(5)  &  0.72(7)    & 0.17   \\
\enddata
\tablenotetext{ a}{ Intrinsic line strength}
\tablenotetext{ b}{ Dipole moment}
\tablenotetext{ c}{ Brightness temperature of the main hyperfine component.
  Line widths and peak velocities are derived from the hyperfine fitting method (see text). }
\tablecomments{Observed positions are (-40\arcsec, -40\arcsec) and (40\arcsec, 40\arcsec) 
  for CB130-3 and L673-SMM4, respectively. \\
  The numbers in parenthesis represent three times the standard deviation in the Gaussian fit 
  in units of the last significant digits. }
\tablerefs{1: \citet{murakami1990}; 2: \citet{yamamoto1990}; 
 3: \citet{lafferty1978}; 4: \citet{alexander1976}; 
 5: \citet{haese1979}; 6: \citet{blackman1976}; 7: \citet{havenith1990}}
\end{deluxetable}

\begin{deluxetable}{llllllll}
\tabletypesize{\scriptsize}
\tablewidth{0pt}
\tablecaption{Parameters for the NH$_{3}$ lines observed with the MPIfR 100~m telescope 
\label{tab-obsnh3}}
\tablehead{
\colhead{Source} & 
  \colhead{$T_{B}$\tablenotemark{a}} & \colhead{$v_{lsr}$} &
  \colhead{$\Delta v$} &  \colhead{} & 
  \colhead{$N$[NH$_{3}$]} & \colhead{$T_{ex}$} &   \colhead{$T_{rms}$} \vspace{2mm} \\
\colhead{Name} & 
  \colhead{(K)} & \colhead{(km s$^{-1}$)} &
  \colhead{(km s$^{-1}$)} &  \colhead{$\tau_{\rm{main}}$\tablenotemark{b}} & 
  \colhead{(10$^{14}$ cm$^{-3}$)} & \colhead{(K)} &   \colhead{(K)} 
}
\startdata
CB130-3\tablenotemark{c}     &   1.07  &  7.26(3) &  0.78(6)  &  0.3    & 1.63($^{+125}_{-26}$)    &  6.5      & 0.14 \\
L673-SMM4                    &   2.8   &  6.72(1) &  0.43(3)  &  1.8(4) & 5.1(14)                  &  6.6(15)  & 0.11 \\
L492\tablenotemark{d}        &   2.5   &  7.81(1) &  0.30(2)  &  3.5(5) & 5.5(12)                  &  5.6(10)  & 0.13 \\
TMC-1\tablenotemark{d}       &   1.91  &  5.99(2) &  0.50(4)  &  1.1(5) & 3.2(17)                  &  6.2(28)  & 0.16 \\
\enddata
\tablenotetext{ a}{Peak brightness temperature of the main hyperfine component.}
\tablenotetext{ b}{Optical depth of the main hyperfine components.}
\tablenotetext{ c}{Hyperfine fitting was not successful. 
Only the main hyperfine component was fitted to the Gaussian profile, 
and the column density was derived by assuming the excitation temperature of 6.5~K (see text). }
\tablenotetext{ d}{See Appendix.}
\tablecomments{The numbers in parenthesis represent three times the standard deviation in the Gaussian fit 
in units of the last significant digits.}
\end{deluxetable}

\begin{deluxetable}{lll}
\tabletypesize{\scriptsize}
\tablewidth{0pt}
\tablecaption{Column densities toward CB130-3 and L673-SMM4
\label{tab-collte}
}
\tablehead{
\colhead{Molecule} & \colhead{CB130-3} & \colhead{L673-SMM4}}
\startdata
CCS              &  2.2$^{+0.6}_{-0.2}$       &  3.8$^{+0.9}_{-0.4}$       \\
HC$_{3}$N        &  2.8$^{+1.0}_{-0.4}$       &  6.8$^{+5.8}_{-1.5}$       \\
HC$_{5}$N        &  0.81$^{+0.22}_{-0.15}$    &  4.3$^{+1.3}_{-0.8}$       \\ 
H$^{13}$CO$^{+}$ &  0.072$^{+0.034}_{-0.018}$ &  0.123$^{+0.055}_{-0.030}$ \\
DNC              &  0.20$^{+0.16}_{-0.12}$    &  0.74$^{+0.65}_{-0.46}$    \\
HN$^{13}$C       &  0.156$^{+0.13}_{-0.10}$   &  0.38$^{+0.34}_{-0.25}$    \\
N$_{2}$H$^{+}$   &  0.22$^{+0.04}_{-0.03}$    &  1.21$^{+0.20}_{-0.19}$    \\
NH$_{3}$         & 16.3$^{+12.5}_{-2.6}$      & 51$^{+14}_{-14}$           \\
\enddata
\tablecomments{The column densities are in unit of 10$^{13}$~cm$^{-2}$. }
\end{deluxetable}

\begin{deluxetable}{lccccccccc}
\tabletypesize{\scriptsize}
\tablewidth{0pt}
\tablecaption{Molecular abundance ratios 
\label{tab-dhratio}}
\tablehead{
\colhead{Molecule} & 
  \colhead{L1521E} &  \colhead{L1521B} &  \colhead{L1495B} & \colhead{TMC-1(CP)} & 
  \colhead{CB130-3} & \colhead{L492}   &  \colhead{L673-SMM4} & \colhead{L1544} & \colhead{L1498}   
}
\startdata
DNC/HN$^{13}$C & 0.66 & 0.70 & $<$0.66 & 1.25 & 1.28$^{+0.27}_{-0.05}$ &  1.27 &  1.96$^{+0.32}_{-0.01}$ &  3.0 &  1.91  \\
NH$_{3}$/CCS   & 2.6  & 3.5  &    3.8  & 4.8  & 7.6$^{+3.0}_{-0.4}$    & 10.4  & 13.6$^{+0.3}_{-2.5}$    & 15   & 25     \\
\enddata 
\tablecomments{For CB130-3 and L673-SMM4, the results of the present study are given. 
For the other sources, the values are taken from \citet{hirota2006} and references therein, 
except for the NH$_{3}$/CCS ratios in TMC-1(CP) and L492 (see Appendix). }
\end{deluxetable}

\begin{figure} \begin{center}
\includegraphics[width=7.5cm]{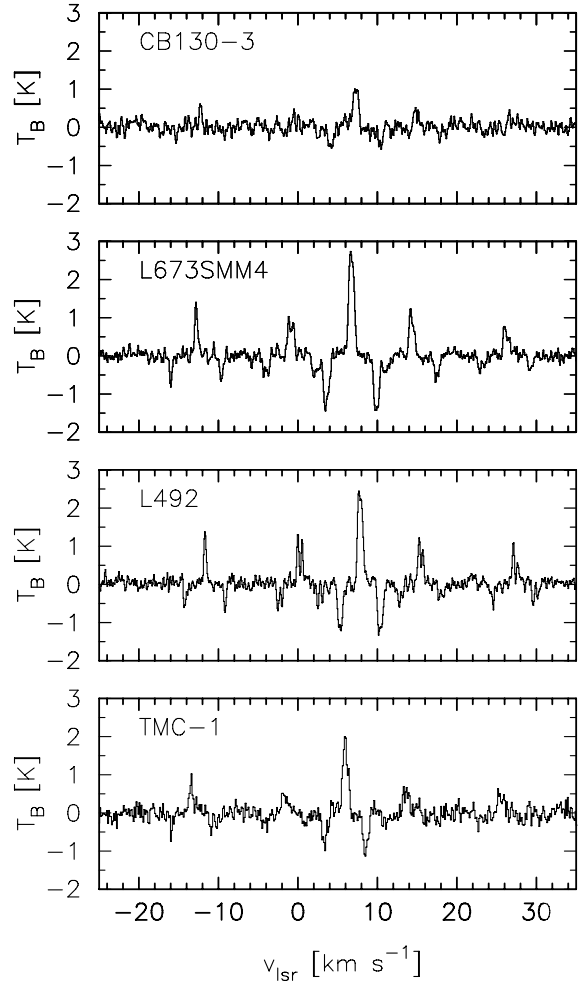}
\caption{Spectra of the NH$_{3}$ (1,1) lines. 
The absorption-like features below the baseline 
are due to artifacts of the frequency switching observations. 
The results for L492 and TMC-1 are also plotted (see Appendix). 
\label{fig-nh3spectra}  }
\end{center} \end{figure}

\begin{figure} \begin{center}
\includegraphics[width=7.5cm]{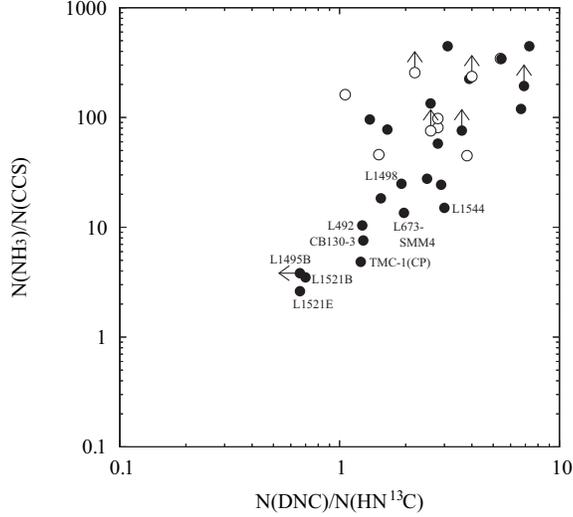}
\caption{Relationship between the abundance ratios of DNC/HN$^{13}$C and 
NH$_{3}$/CCS.  
Filled and open symbols represent the abundance ratios for starless and 
star-forming cores, respectively. The data were originally presented 
in \citet{hirota2001}. We have revised the data for CCPRs 
\citep{hirota2002, hirota2003, hirota2004, hirota2006}. 
\label{fig-plots} }
\end{center} \end{figure}

\appendix
\section{Follow-up Observations of the NH$_{3}$ Lines toward L492 and TMC-1}

In Figure \ref{fig-plots}, we plot the 
relationship between the DNC/HN$^{13}$C and NH$_{3}$/CCS ratios. 
The original plot was reported by \citet{hirota2001} and we have here revised it 
by employing the present results for CB130-3 and L673-SMM4. 
In addition, we have also involved the result of the L492 observation in the present study. 
For L492, the observed position of the NH$_{3}$ line was 
shifted from the CCS peak position \citep{hirota2006}. 
Therefore, we have employed the NH$_{3}$ data toward the CCS peak position 
(40\arcsec, 0\arcsec) of L492 observed with the MPIfR 100~m telescope 
as described in section \ref{sec-observations}.  
In addition, we have also replaced the NH$_{3}$ data for the cyanopolyyne peak of 
TMC-1 with that obtained with the MPIfR 100~ m telescope. 
The result of the NH$_{3}$ observation is summarized in Table \ref{tab-obsnh3} 
and Figure \ref{fig-nh3spectra}. 
The NH$_{3}$ column densities in L492 and TMC-1 become larger by a factor of 
1.6 and 1.7, respectively. 
For the DNC/HN$^{13}$C ratios, we have 
employed recently observed data reported by \citet{hirota2002}, 
\citet{hirota2003}, \citet{hirota2004}, \citet{hirota2006}, and the present study. 

\end{document}